\documentclass[aps,showpacs,twocolumn,showkeys,amsmath,amssymb,floatfix,nofootinbib,preprintnumbers]
{revtex4}
\pdfoutput=1
\usepackage{graphicx}
\usepackage{bm}
\usepackage{hyperref}
\usepackage{epstopdf}
\usepackage{float}

\begin{document}

\title{Tetraquark Cusp Effects from Diquark Pair Production}

\author{Samuel H. Blitz}
\email{sblitz@asu.edu}

\author{Richard F. Lebed}
\email{richard.lebed@asu.edu}
\affiliation{Department of Physics, Arizona State University, Tempe,
Arizona 85287-1504, USA}

\date{March, 2015}

\begin{abstract}
  We present the first study of the cusp effect (the movement of
  resonant poles due to the proximity of multiparticle thresholds)
  caused by the creation of diquark-antidiquark pairs.  The cusp
  profile for such states is obtained from constituent counting rules.
  We compare the effectiveness of diquark cusps in moving resonant
  poles with that from a phenomenological form commonly used for
  meson-pair creation, and find that mesons tend to be more effective
  at lower energies $({\it e.g.}$, the $K\overline{K}$ threshold),
  while diquarks tend to be more effective at the charm [$X(3872)$]
  scales and above.
\end{abstract}


\maketitle


\section{Introduction} \label{sec:Intro}

One of the most important discoveries of the past year in hadron
physics was the experimental confirmation by LHCb~\cite{Aaij:2014jqa}
of a new type of hadron joining the $\bar q q$ mesons and $qqq$
baryons -- the {\em tetraquark\/} ($\bar q q \bar q q$) resonance
$Z_c^+ (4430)$.  In fact, this state is just one of an ever-growing
menagerie of hadrons now believed to be tetraquarks, all of which have
been observed since the 2003 discovery by Belle~\cite{Choi:2003ue} of
an unusually narrow charmonium-like state, the $X(3872)$.  However,
until the discovery of charged states (now called $Z_c$), one could
not be certain that the $X(3872)$ was not just an unusual conventional
charmonium state or a hybrid $\bar c c g$; and until the observation
of a phase $\delta$ increase by $\frac \pi 2$ radians in the complex
scattering amplitude in which $Z_c$ is produced~\cite{Aaij:2014jqa},
one could not be certain that the states are true resonances as
opposed to, say, kinematical reflections of $t$-channel exchanges.
There now seems to be little doubt that the $Z_c^+ (4430)$ is a
tetraquark resonance with $J^P = 1^+$ quantum numbers and valence
quark structure $\bar c c \bar d u$. In addition, all indications
suggest that the charge-zero $X(3872)$ is a $J^P = 1^+$ $\bar c c \bar
q q$ state, where $\bar q q$ is some linear combination of $\bar u u$
and $\bar d d$.

But waiting for the final confirmation of the tetraquark label did not
preclude early speculation on the composition and structure of such
states.  The initial interpretation -- and still the most prevalent
one -- is that the tetraquarks are di-meson molecules, bound together
by color van der Waals-type forces (see~\cite{Brambilla:2014aaa} for a
recent review). This interpretation is suggested by the proximity of
the mass of several of the states to the corresponding two-meson
thresholds.  For example, $m_{X(3872)} - m_{D^{*0}} - m_{D^0} = -0.11
\pm 0.21$~MeV\@.  However, several other tetraquark candidates lie
just above such thresholds -- clearly muddling the simple bound-state
interpretation -- and several others lie nowhere near any two-meson
thresholds.  Furthermore, the substantial prompt production cross
section for the $X(3872)$ at colliders seems to be incompatible with
the state being solely composed of loosely-bound meson
pairs~\cite{Esposito:2013ada,Guerrieri:2014gfa} -- even taking into
account a substantial modification due to final-state
interactions~\cite{Artoisenet:2009wk,Artoisenet:2010uu}.  Another
well-known interpretation is {\it
  hadro-charmonium\/}~\cite{Voloshin:2007dx}, in which an ordinary
charmonium state lies at the core of a light-quark cloud, although the
cohesiveness of such states, and the extent to which they mix with
conventional charmonium states, is unclear.

In this work, we employ yet another noteworthy interpretation for the
tetraquark states, that of a diquark-antidiquark ($\delta$-$\bar
\delta$) bound-state pair.  Originally proposed for charmonium-like
tetraquarks in Ref.~\cite{Maiani:2004vq}, the diquark picture has the
advantage of possessing a much richer color dynamics (the diquarks
necessarily being color nonsinglets), but it also has the potential
drawback of producing many more tetraquark states than are currently
observed.  However, if one assumes that the dominant interactions are
due to spin-spin couplings within each diquark, one can obtain a rather
satisfactory accounting of the presently known tetraquark
candidates~\cite{Maiani:2014aja}.

One may wonder why the component quarks in a $\delta$-$\bar \delta$
bound state do not immediately rearrange themselves into color-singlet
$\bar q q$ pairs, returning one to the molecular picture.  Indeed, the
strength of the color force between quarks (or antiquarks) in
SU(3)-color representations $R_1$ and $R_2$ coupling to a
representation $R$ is proportional to the combination of quadratic
Casimirs given by $C_2 (R) - C_2(R_1) - C_2 (R_2)$.  The only
attractive channels are the $\bar q q$ singlet ($R_1 = \bar {\bf 3}$,
$R_2 = {\bf 3}$, $R = {\bf 1}$) and the $q q$ antitriplet ($R_1 = R_2
= {\bf 3}$, $R = \bar {\bf 3}$), with the former being twice as strong
as the latter.

Rather than treating the tetraquark as a metastable $\delta$-$\bar
\delta$ bound state, we proposed in Ref.~\cite{Brodsky:2014xia} a new
paradigm, in which the tetraquarks are the quantized modes of a color
flux-tube stretched between a rapidly separating $\delta$ (color-$\bar
{\bf 3}$) and $\bar \delta$ (color-{\bf 3}) pair.  This picture
naturally explains why many, but not all, of the tetraquarks lie near
hadron thresholds (energies at which the color string is allowed to
break); for example, the state $X(4632)$ lies only slightly higher
than the lightest charmed-baryon $\Lambda_c^+ \bar \Lambda_c^-$
threshold at 4573~MeV, and it decays dominantly to this baryon pair,
as is explained by the fragmentation of the flux-tube to a light $\bar
q q$ pair.  The widths of the states below this threshold are not
especially large, because they hadronize by wave function overlaps
with the mesons formed from quarks in $\bar \delta$ and antiquarks in
the $\delta$, which in turn can achieve a substantial separation (over
1~fm) if sufficient initial kinetic energy is imparted to the system,
as in a $B$-meson decay.  Evidence for this large separation is
apparent in the decay of the $Z_c (4430)$, which greatly favors
coupling to $\psi (2S) \, \pi$ rather than $J/\psi \,
\pi$~\cite{Yuan:2014rta} -- even though the charmonium states have the
same $J^{PC} = 1^{--}$ quantum numbers and there is smaller phase
space for the heavier (and spatially much larger) $\psi(2S)$.

This paper presents the first dynamical study of the $\delta$-$\bar
\delta$ tetraquark picture, using the well-known {\it constituent
  counting rules\/}~\cite{Brodsky:1973kr,Matveev:1973ra,
  Brodsky:1974vy,Farrar:1979aw,Efremov:1978rn,Duncan:1979hi,
  Duncan:1980qd,Lepage:1980fj,Sivers:1982wk,Mueller:1981sg,
  Brodsky:1989pv}, which determine the fall-off scaling of hard
exclusive processes in the high-energy regime.  In essence, the
counting rules predict that the cross sections, or equivalently, the
form factors for processes at high values of Mandelstam $s$ at fixed
$\theta_{\rm cm}$, fall off as a power of $s$ directly determined by
the total number (incoming plus outgoing) of fundamental constituents
participating in the hard subprocess.  Here we are interested in
comparing the effects of $\delta$-$\bar \delta$ state production with
that due to meson-meson states on the masses of tetraquark resonances
coupled to them, using dispersion relation techniques.  In another
paper to appear~\cite{Brodsky:2015XXX}, we will directly discuss the
phenomenological uses of the cross-section scalings themselves.

The method to be used is the well-known {\it cusp effect}, in which
the presence of thresholds for the opening of on-shell states coupled
to resonances creates a modification to the self-energy function that
tends to drag the bare resonant pole mass toward the threshold.  The
basic idea appears to have been known since the early 1960's, but was
first presented in its current form in the mid
1990's~\cite{Tornqvist:1995kr}, and first applied to the new heavy
exotics in 2008~\cite{Bugg:2008wu}.  Furthermore, recent
calculations~\cite{Guo:2014iya} show that these states cannot {\em
  just\/} be cusps (although ones in the $B\bar B$ system might
be~\cite{Bugg:2011jr,Swanson:2014tra}); the presence of real
resonances is required.  In this paper we compare the effectiveness of
this dragging effect due to the presence of both $\delta$-$\bar
\delta$ and two-meson thresholds, and find first, that the potential
amount of shifting of resonant poles decreases for heavier-quark
systems ($K\overline{K}$ vs.\ $D\bar D^*$ vs.\ $B\bar B^*$), and
second, that the $\delta$-$\bar \delta$ states become more effective
at pole-dragging than two-meson states for the $D\bar D^*$ and $B\bar
B^*$ thresholds associated with the new heavy exotic resonances.
Since $\delta$ is not a color singlet, using a $\delta$-$\bar \delta$
threshold in a QCD dispersion relation (where ``on shell'' usually
means not only that the particles are not virtual, but also
asymptotically free) must be interpreted with some care; for the
purposes of this calculation, we assert that the substantial
$\delta$-$\bar \delta$ separation advocated in \cite{Brodsky:2014xia}
creates states that are in a sense ``almost'' free, and therefore
possess an on-shell threshold.

This paper is organized as follows: In Sec.~\ref{sec:Cusp} we briefly
review the meaning and origin of the cusp effect, and establish the
mathematical formalism used for its implementation, with some details
relegated to the Appendix.  In Sec.~\ref{sec:DiqThresh} we address in
greater detail the issue of whether diquark pairs truly produce
physically meaningful thresholds.  Section~\ref{sec:QuarkCount}
presents a brief overview of the constituent counting rules used to
establish the large-energy scaling of the $\delta$-$\bar \delta$ form
factor.  In Sec.~\ref{sec:Cases} we describe the algebraic results for
both the mesonic and diquark forms, and present our numerical results
in Sec.~\ref{sec:Results}.  In Sec.~\ref{sec:Concl} we summarize and
indicate future directions.

\section{The Cusp Effect} \label{sec:Cusp}

The cusp effect has a rather straightforward origin in the analyticity
of the self-energy functions $\Pi(s)$ that appear in the propagator of
resonant or bound states and source the creation of their decay
products.  Closely following notation introduced in
Refs.~\cite{Tornqvist:1995kr,Bugg:2008wu}, we start with the
propagator denominator
\begin{equation} \label{eq:Prop}
P^{-1}_{\alpha \beta} (s) = (M_{0,\alpha}^2 - s) \delta_{\alpha
\beta} - \Pi_{\alpha \beta} \, ,
\end{equation}
where $\alpha, \beta$ index resonances that can mix; for simplicity,
in this work we assume only unmixed propagating states and henceforth
suppress this index.  The sign of $\Pi(s)$ is chosen to match that
appearing in the majority of quantum field theory texts ({\it e.g.},
\cite{Peskin:1995ev}); when positive, it is seen to correspond to an
attractive interaction.

$\Pi (s)$ is analytic everywhere in the complex $s$ plane except for
cuts (or poles) along the positive real $s$ axis that result from the
opening of on-shell channels, which we label by $i$.  Consider
explicitly the creation of such two-particle states of masses
$m_{1,i}$, $m_{2,i}$ via the form factors $F_i(s)$, conventionally
normalized to unity at threshold, $s_{{\rm th}, i} \equiv (m_{1,i} +
m_{2,i})^2$.  Defining the (dimensionful) coupling constant to state
$i$ as $g_i$, one has
\begin{equation} \label{eq:ImPi}
{\rm Im} \,\Pi (s) = \sum_i g_i^2 \rho_i (s) \, F_i^2 (s) \, \theta
( s - s_{{\rm th}, i} ) \, ,
\end{equation}
and $\rho_i$ is the two-body phase-space factor given in terms of the
c.m.\ momentum $k_i$ or the K\"{a}ll\'{e}n function $\lambda$ by
\begin{equation} \label{eq:rhodef}
\rho_i (s) \equiv \frac{2k_i}{\sqrt{s}} = \frac{\lambda^{\frac 1 2}
( s, m_{1,i}^2, m_{2,i}^2 )}{s} \, .
\end{equation}
Defining $s_{{\rm th}, 1}$ as the lowest threshold (and therefore the
branch point whose cut extending to $s = \infty$ overlaps all others),
one may apply Cauchy's theorem to a contour that goes around this cut,
obtaining the standard unsubtracted dispersion relation:
\begin{equation} \label{eq:unsubtracted}
{\rm Re} \, \Pi (s) = \frac 1 \pi \, {\rm P} \!
\int_{\! s_{{\rm th}, \, 1}}^\infty ds^\prime \,\frac{{\rm Im} \, \Pi
(s^\prime)}{s^\prime - s} \, ,
\end{equation}
where P indicates the standard Cauchy principal value prescription.
For reasons to be discussed below, we find it useful to use the
once-subtracted form (at $s=0$) of Eq.~(\ref{eq:unsubtracted}):
\begin{equation} \label{eq:oncesubtracted}
{\rm Re} \, \Pi (s) =  {\rm Re} \, \Pi (0) + \frac{s}{\pi} \, {\rm P}
\! \int_{\! s_{{\rm th}, \, 1}}^\infty \frac{ds^\prime}{s^\prime} \,
\frac{{\rm Im} \, \Pi (s^\prime)}{s^\prime - s} \, ,
\end{equation}

One notes immediately from Eqs.~(\ref{eq:ImPi})--(\ref{eq:rhodef})
that Im~$\Pi(s)$ is zero on the real $s$ axis for $s < s_{{\rm th},
  1}$, and positive for $s > s_{{\rm th}, 1}$.  One further sees from
Eq.~(\ref{eq:unsubtracted}) or Eq.~(\ref{eq:oncesubtracted}) that the
integrand in Re~$\! \Pi(s)$ is positive for all $s < s_{{\rm th} , 1}$
since $s^\prime - s > 0$ in the entire integration range, and
therefore acts as an attractive interaction.  As shown below, explicit
functional forms for $F(s)$ give rise to a positive Re~$\! \Pi(s)$
that reaches a peak, or {\em cusp}, at $s_{{\rm th}, 1}$ and falls off
in either direction, remaining always positive for $s < s_{{\rm th},
  1}$ but generally passing through the axis (and effectively creating
a potential barrier) for some value of $s > s_{{\rm th}, 1}$. This
effective attraction acts to pull the pole position of the propagator
Eq.~(\ref{eq:Prop}) toward $s_{{\rm th}, 1}$, a synchronization of the
resonance with a threshold (although in Sec.~\ref{sec:Results} we see
some interesting exceptions to this expectation).

In the bulk of this work, for simplicity we specialize to the
equal-mass case $m_{1,i} = m_{2,i} \equiv m_i$, which is rather
closely satisfied for the cases of experimental interest.  Full
expressions analogous to the ones appearing in the text below are
presented in the Appendix, where it is seen that the relevant
expansion parameter is
\begin{equation} \label{eq:epsilondef}
\epsilon \equiv \left( \frac{m_{1,i} - m_{2,i}}{m_{1,i} + m_{2,i}}
\right)^2 \, .
\end{equation}
For $X(3872)$, with $m_{1,i} = m_{D^0}$ and $m_{2,i} = m_{\bar
  D^{*0}}$, one finds $\epsilon = 1.35 \cdot 10^{-3}$.  In the
equal-mass case,
\begin{equation} \label{eq:rhodef2}
\rho_i (s) = \sqrt{ 1 - \frac{4m_i^2}{s} }
= \sqrt{ 1 - \frac{s_{{\rm th} i}}{s} } \, ,
\end{equation}
and its analytic continuation for $s < s_{{\rm th}, i} = 4m_i^2$ is
\begin{equation} \label{eq:vdef}
v_i (s) = \sqrt{ \frac{4m_i^2}{s} - 1}
= \sqrt{ \frac{s_{{\rm th},i}}{s} - 1} \, .
\end{equation}

\begin{figure}
\begin{center}
\includegraphics[width=\linewidth]{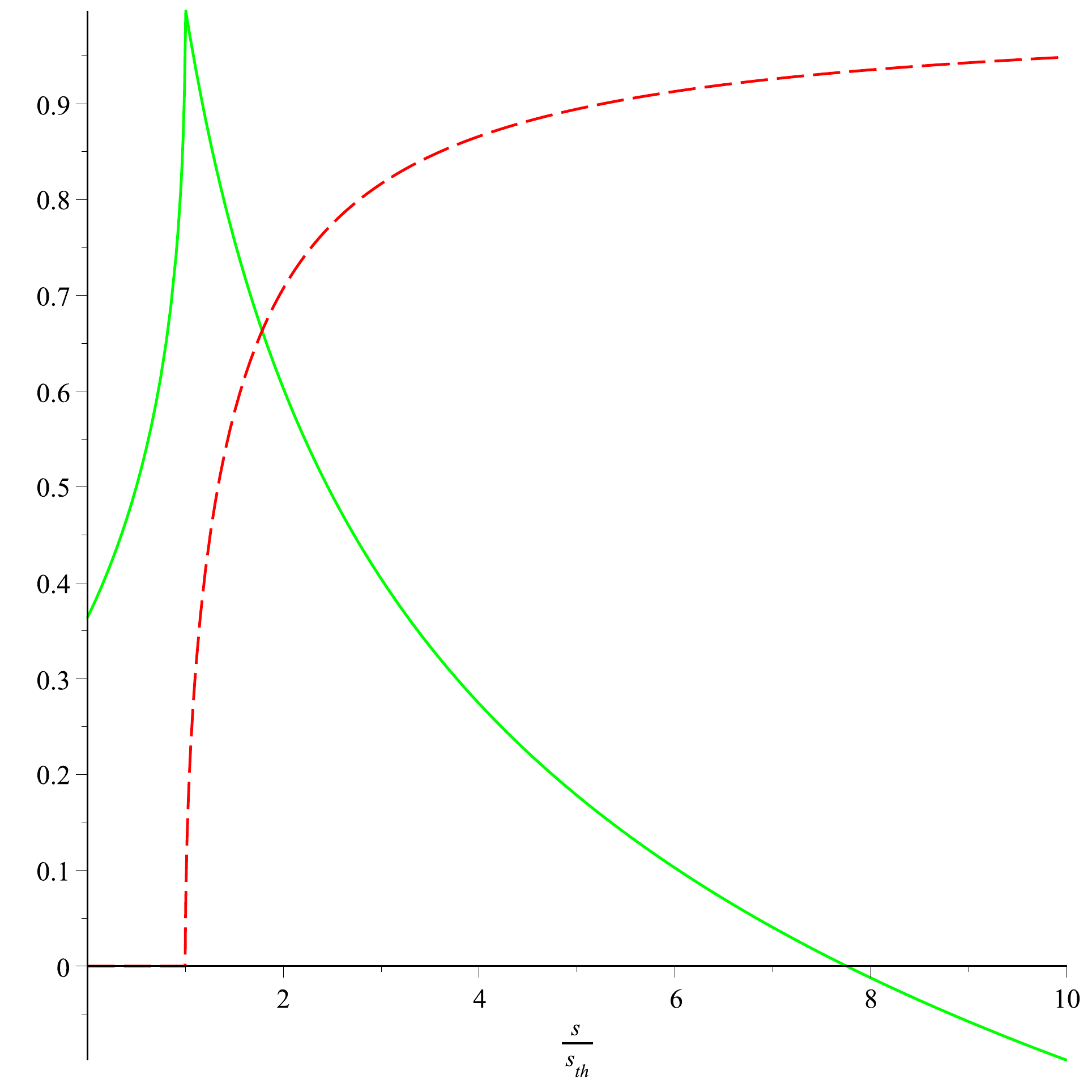}
  \caption{The threshold cusp in Re~$\! \Pi_i(s)/g_i^2$ (green, solid)
    and corresponding Im~$\! \Pi_i(s)/g_i^2$ (red, dashed) for form
    factor $F_i(s)=1$, as given by Eq.~(\ref{eq:F=1cusp}).  $s$ is
    expressed in units of $s/s_{{\rm th},i}$, and Re~$\! \Pi_i(s)/g_i^2$
    is shifted to equal 1 at $s = s_{{\rm th},i}$.
    \label{Fig:F=1cusp}}
\end{center}
\end{figure}

To illustrate the cusp explicitly, let us consider the special case
$F_i (s) = 1$ described in~\cite{Bugg:2008wu}.  This particular choice
is unphysical because unitarity and perturbative QCD require real
hadronic form factors to fall off at large $s$, but it is an
interesting toy model for cases in which the dispersion relation is
dominated by the form factor near threshold.  It is also interesting
because the dispersion integrals can be carried out explicitly; the
nonvanishing large-$s$ behavior of $F_i(s)$ requires one to use the
once-subtracted form Eq.~(\ref{eq:oncesubtracted}) to obtain finite
results at finite $s$.  Allowing the constant Re~$\! \Pi_i(0) = -\frac
2 \pi$ to renormalize the pole mass $M_0$ in the propagator
Eq.~(\ref{eq:Prop}), one finds:
\begin{eqnarray} \label{eq:F=1cusp}
\frac{{\rm Re} \, \Pi_i (s)}{g_i^2} & = & \frac{\rho_i}{\pi} \ln
\frac{1-\rho_i}{1+\rho_i} = -\frac{2\rho_i}{\pi} \tanh^{-1} \! \rho_i
\, , \ \ s \geq s_{{\rm th}, i} \, , \nonumber \\
& = & -v_i + \frac{2v_i}{\pi} \tan^{-1} v_i \, , \ \ s < s_{{\rm th},
i} \, .
\end{eqnarray}
Using Eqs.~(\ref{eq:rhodef2})--(\ref{eq:vdef}), note that the
tan$^{-1}$ and tanh$^{-1}$ terms are analytic continuations of the
same function of $s$ in different regimes; the only unmatched term is
$-v_i$, which is induced by analyticity of $\Pi(s)$ and the
discontinuity of Im~$\! \Pi_i(s_{{\rm th}, i})$.  In
Fig.~\ref{Fig:F=1cusp}, we plot Eq.~(\ref{eq:F=1cusp}) [and Im~$\!
\Pi_i(s_{{\rm th}, i})/g_i^2$] as a function of $s/s_{{\rm th},i}$,
and see that the below-threshold term $-v_i$ is responsible for the
cusp.

Cusp behavior also appears for more realistic choices of form factor
$F_i(s)$, although the dispersion relation integrations must often be
carried out numerically.  Two features of the integrals hamper the
convergence of numerical calculations: First, the integration range
stretches to $s = \infty$, and second, the denominator factor
$s^\prime \! - s$ produces a logarithmic singularity when $s > s_{{\rm
    th},i}$.  A simple change of variable cures the first problem;
noting from Eq.~(\ref{eq:rhodef2}) that $s \in [s_{{\rm th}, i},
\infty)$ maps to $\rho_i \in [0,1)$, one defines the integration
variable $\rho_i^\prime$ analogously to Eq.~(\ref{eq:rhodef2}), but
with $s \to s^\prime$.  Then one finds the unsubtracted relation
Eq.~(\ref{eq:unsubtracted}) to become
\begin{equation}
\frac{{\rm Re} \, \Pi_i(s)}{g_i^2} = \frac{2}{\pi} \frac{s_{{\rm
th},i}}{s} \, {\rm P} \! \int_0^1 d\rho^\prime_i
\frac{\rho^{\prime \, 2}_i}{1-\rho^{\prime \, 2}_i}
\frac{1}{\rho^{\prime \, 2}_i - \rho^2_i} F_i^2 \left( s^\prime
\right) \, ,
\end{equation}
where [inverting Eq.~(\ref{eq:rhodef2})] $s^\prime = s_{{\rm th},i} /
(1-\rho^{\prime \, 2}_i)$.  In fact, the once-subtracted relation
Eq.~(\ref{eq:oncesubtracted}) becomes even simpler:
\begin{equation} \label{eq:rhodispreln}
\frac{1}{g_i^2} \left[ {\rm Re} \, \Pi_i (s) - {\rm Re} \, \Pi_i (0)
\right] = \frac{2}{\pi} \, {\rm P} \! \int_0^1 d\rho^\prime_i
\frac{\rho^{\prime \, 2}_i}{\rho^{\prime \, 2}_i - \rho^2_i} F_i^2
\left( s^\prime \right) \, .
\end{equation}
Since, as discussed above, one expects Re~$\! \Pi_i (s) \to 0$ as $s
\to \infty$ for physical $F_i(s)$, no information is lost by using the
once-subtracted and functionally simpler form
Eq.~(\ref{eq:rhodispreln}).  The logarithmic singularity does not
arise for $s < s_{{\rm th}, i}$; since $\rho_i^2 = -v_i^2$ according
to Eqs.~(\ref{eq:rhodef2})--(\ref{eq:vdef}), one may rewrite the
integrand factor in Eq.~(\ref{eq:rhodispreln}) for $s < s_{{\rm th},
  i}$ as the nonsingular form
\begin{equation}
\frac{\rho^{\prime \, 2}_i}{\rho^{\prime \, 2}_i - \rho^2_i} \to 1 -
\frac{v^2_i}{\rho^{\prime \, 2}_i + v^2_i} \, .
\end{equation}
The logarithmic singularity for $s \ge s_{{\rm th}, i}$ is readily
handled via an integration by parts on the $1/(s^\prime - s)$ factor
in Eq.~(\ref{eq:oncesubtracted}), followed by conversion to the
integration variable $\rho^\prime$.  One obtains:
\begin{eqnarray}
\lefteqn{\frac{1}{g_i^2} \left[ {\rm Re} \, \Pi_i (s) - {\rm Re} \,
\Pi_i (0) \right] = } & & \nonumber \\
& & \hspace{-1em} -\frac{1}{\pi} \int_0^1 d\rho^\prime_i \left[ F^2_i
(s^\prime) + \frac{2 s_{{\rm th},i} \, \rho^{\prime \,
2}_i}{(1 - \rho^{\prime \, 2}_i)^2} \frac{dF^2_i(s^\prime)}{ds^\prime}
\right] \ln \frac{\left| \rho^{\prime \, 2}_i - \rho^2_i \! \right|}
{ 1 - \rho_i^2 } \, . \nonumber \\ & & \label{eq:rhodispreln2}
\end{eqnarray}
The singularity at $\rho_i = \rho^\prime_i$ remains, but now it is
integrable for any smooth $F_i^2(s)$ that asymptotes to zero as $s \to
\infty$.  Moreover, the non-logarithmic part of this integral is just
the exact differential $d[\rho_i(s^\prime) F_i^2 (s^\prime)]$, whose
argument vanishes at $s^\prime = s_{{\rm th},i}$ and also at $s^\prime
= \infty$ if $F_i^2 (s^\prime)$ falls off faster than
$1/\rho_i(s^\prime)$; in such a case, the
($\rho_i^\prime$-independent) term $1 - \rho_i^2$ in the logarithm may
also be deleted.  Since the function obtained from
Eq.~(\ref{eq:oncesubtracted}), {\it i.e.}, the explicit left-hand side
of (\ref{eq:rhodispreln}) or (\ref{eq:rhodispreln2}), is central to
our analysis, let us henceforth abbreviate it as $\pi_i (s) \equiv [
{\rm Re} \, \Pi_i (s) - {\rm Re} \, \Pi_i (0) ] / g_i^2$.

\section{Do Diquark Pairs Have Thresholds?} \label{sec:DiqThresh}

Before proceeding, one must address the issue mentioned in the
Introduction of whether a $\delta$ state, being colored, really
possesses a well-defined mass, and therefore whether the
$\delta$-$\bar \delta$ state really possesses a well-defined
threshold.  Strictly speaking, the quoted masses of colored particles
like quarks depend upon the choice of renormalization scheme, and
therefore do not carry the same status of being observables as do
meson masses.  Nevertheless, quark masses are extracted using a
variety of methods (quark models, lattice gauge theory, sum rules,
{\it etc.}), and the numerical values thus obtained are bestowed with
some degree of physical significance.

An elegant discussion of such issues appears in the classic text by
Georgi~\cite{Georgi:1985kw} (Sec.~3.2); he considers a universe in
which the strong coupling constant $\alpha_s$ is as small as
$\alpha_{\rm EM}$, but confinement still occurs.  The scale
$\Lambda_{\rm QCD}$ then becomes extremely large ($\simeq
10^{20}$~cm), and for all practical purposes, quarks would be just as
visible in experiments as electrons, and would possess easily
determined quantum numbers such as mass.  Georgi then continues the
analysis by increasing $\Lambda_{\rm QCD}$ towards its physical
$O(250~{\rm MeV})$ value, and argues that not until $\Lambda_{\rm QCD}
\simeq \alpha_s m_q$ (including the running value of $\alpha_s$) do
quarks become inseparable from their hadrons, and hence hidden from
view.

We take this lesson as our touchstone, that substantial physical
separation implies identifiability as particles with
well-characterized masses, even in the presence of confinement.  The
color dynamics allows diquarks to form as bound states, and kinematics
permits them to separate a distance before being forced to hadronize.
The salient question then becomes: How far is far enough, before the
diquarks can reasonably be said to appear as identifiable particles
possessing a well-defined mass threshold?  We argue that this event
occurs when the $\delta$-$\bar \delta$ state can no longer be mistaken
for one in which the diquark wave functions have substantial wave
function overlap.  Of course, the size of diquark wave functions is
not at all a known quantity, but we can obtain a reasonable estimate
by considering the size of mesons with analogous quark content. That
is, instead of $cu$ and $\bar c \bar d$ diquarks, we consider $D$
mesons.  Again, meson radii are not directly observable quantities and
the precise nature of the fall-off of their wave functions with $r$ is
unknown, but at least in this case some calculations have been
performed.  For example, Ref.~\cite{Hwang:2001th} calculates the
electromagnetic charge radius of the $D^+$ to be 0.43~fm, rather
smaller than charge radii of the unflavored mesons, due to the
presence of the heavy $c$ quark.  One may expect the diquarks to be
slightly larger because the initial color attraction between the quark
components is smaller; however, the true size is dominated by some
combination of nonperturbative gluodynamics and the heavy quark mass,
so we expect the diquarks to be not much larger. We take from this
result that 1~fm is not an unreasonable estimate for the onset of
significant separation between the diquarks, and hence the
identification of a well-defined diquark mass and pair-production
threshold.  Again, an important ingredient in this estimate appears to
be the presence of heavy quarks, which might very well be related to
the reason why exotics have first become clearly visible in the charm
system.

Furthermore, as first pointed out in Ref.~\cite{Tornqvist:1995kr}, the
cusp effect treatment of the previous section holds at full strength
[as in Eq.~(\ref{eq:rhodef})] only for thresholds of particles in a
relative $s$ wave.  Only certain $\delta$-$\bar \delta$ states will
fall into this category, in precisely the same way that the
conventional cusp effect is most prominent for thresholds of mesons
produced in a relative $s$ wave.

A very interesting question is how the diquark thresholds disappear if
their separation is insufficient to prevent a large wave function
overlap.  We speculate that these thresholds gradually dissolve into
correlations between the couplings to meson pairs to which the exotics
preferentially decay, and such correlations would create a smeared
``cusp'' that would gradually disappear for smaller and smaller
diquark separations.  The exact separation at which such a structure
first appears, and hence the assumption that a viable diquark cusp
effect is supportable, is very much open to debate, but we believe
1~fm is a reasonable estimate.  The diquark cusps studied in this
paper are of course based upon an idealized picture of this behavior.

\section{Constituent Counting Rules} \label{sec:QuarkCount}

In this section we present a brief summary of the constituent counting
rules developed over a number of years following the advent of
perturbative QCD (pQCD)~\cite{Brodsky:1973kr, Matveev:1973ra,
Brodsky:1974vy,Farrar:1979aw,Efremov:1978rn,Duncan:1979hi,
Duncan:1980qd,Lepage:1980fj,Sivers:1982wk,Mueller:1981sg,
Brodsky:1989pv}.  They reflect the underlying conformal features and
scale invariance of the QCD coupling.  More recently, they were
derived nonperturbatively by Polchinski and
Strassler~\cite{Polchinski:2001tt} using AdS/QCD.  A more thorough
pedagogical introduction, particularly with an eye toward discerning
exotic hadronic structure, is presented in
Ref.~\cite{Kawamura:2013iia}; here, we present an abridged discussion
identifying the central points leading to the correct counting.  In
Ref.~\cite{Kawamura:2013yya}, the same authors show how to use the
large-$s$ scaling behavior to study the underlying generalized parton
distributions and distribution amplitudes entering into these
processes.

The constituent counting rules for large-angle scattering processes
({\it i.e.}, those in which no kinematical variable is small due to
the near-collinearity of any particles) at high c.m.\ energy $\sqrt s$
originate from a remarkably simple source: Each of the individual
constituents being scattered must be redirected through a finite angle
by a large momentum transfer.  In this limit, all constituent masses
are negligible, and all three Mandelstam variables $s$, $t$, and $u
\simeq -s-t$ are large, so that one may express all dimensionful
quantities in powers of $s$ and dimensionless coefficients as
functions of $t/s$.  In the pQCD picture, these deflections are
accomplished through hard gluon exchanges; if leptons are included,
then hard electroweak boson exchanges also appear. In the case of
AdS/QCD, the counting rules reflect the twist dimension of the
interpolating field at short distances.

For the moment, let us consider processes in which the constituents
are all fermions.  Then, assuming that each of the $n = n_{\rm in} +
n_{\rm out}$ constituents shares a finite fraction of the total $s$,
the leading-order Feynman diagrams require a minimum of $\frac n 2 -
1$ hard gauge boson exchanges and a minimum of $\frac n 2 - 2$
internal constituent propagators.  These features supply factors of
$1/s^{\frac n 2 -1}$ and $1/\sqrt{s}^{\, (\frac n 2 -1)}$ to the
invariant amplitude ${\cal M}$, respectively.  Noting that each
external constituent fermion field carries a spinor normalization
scaling as $s^{\frac 1 4}$, one sees that all of the fermion scaling
factors cancel except for an overall factor $s$.  In total,
\begin{equation} \label{eq:amplitude}
{\cal M} \propto 1/s^{\frac n 2 - 2} \, .
\end{equation}
Assuming a conventional scattering process in which the constituents
combine into two initial and two final particles, the cross section is
given by
\begin{equation} \label{eq:scaling}
\frac{d \sigma}{d t} = \frac{1}{16 \pi s^2} | {\cal M} |^2 \equiv
\frac{1}{s^{n - 2}} f \left( \frac{t}{s} \right) \, ,\
\end{equation}
where $f$ has the appropriate mass dimension ($M^{2n-8}$) to match
that ($M^{-4}$) of the left-hand side, but does not itself scale as a
power of $s$; its dimensionful factors are essentially the amplitudes
describing the binding of the constituents into the composite states,
{\it i.e.}, decay constants.

Each external gauge boson introduced ({\it e.g.}, turning an
electroproduction process into photoproduction) removes two external
fermion lines [$\sim (\sqrt{s})^2$] and one hard gauge boson
propagator ($\sim 1/s$), leaving the form of the scaling formula for
${\cal M}$ and $d\sigma/dt$ invariant.

Hadronic form factors in the large-$s$ regime can be studied
analogously.  For example, according to Eq.~(\ref{eq:amplitude}), the
electromagnetic (or any other current-produced) form factor $F_{X}(s)$
appearing in the pair-production amplitude ${\cal M}$ for tetraquark
states $X$ scales as
\begin{equation}
F_{X}(s) \to \frac{1}{s^{\frac 1 2 (1+1+4+4) - 2}} =
\frac{1}{s^3} \, .
\end{equation}
While the origin of $1/s$ factors via hard exchanges is natural and
simple, one may worry about technical complications in real
perturbative QCD that disrupt the simple counting.  These effects
include the running of $\alpha_s (s)$, the renormalization-group
scaling of the constituent distribution
amplitudes~\cite{Lepage:1979zb,Efremov:1979qk}, the presence of
Sudakov logarithms~\cite{Sudakov:1954sw,Cornwall:1975ty,Sen:1982bt},
``pinch'' singularities due to the vanishing of internal gluon
propagators~\cite{Landshoff:1974ew}, and endpoint singularities
occurring in configurations where some of the constituents do not
share an $O(1)$ fraction of the hard scale $s$~\cite{Li:1992nu}.
However, it is believed that the net result of these effects is not
sufficiently severe as to change the leading $s$ power scaling for
exclusive processes.

\section{Mesonic vs.\ Diquark Form Factors} \label{sec:Cases}

The discussion of the previous section shows that the pair-production
form factor $F_i (s)$ for a state with $n$ constituents in the
high-$s$ regime scales as $1/s^{n-1}$.  Of course, this scaling is not
expected to hold in the low-$s$ region, particularly since confinement
physics is not taken into account in this fundamentally perturbative
approach.

\subsection{Meson Form Factors}

In the case of pair creation of conventional mesons, an exponential
form is traditionally favored by phenomenological fits to data.  For
example, for $K \bar K$ production, Ref.~\cite{Bugg:2008wu} uses
$F^2_i= {\rm exp} (-k^2_i R^2/3)$, where $R = 0.6$~fm, while for $B
\bar B^*$ production, Ref.~\cite{Swanson:2014tra} uses $F^2_i = {\rm
  exp} (-s/\beta_i^2)$ (once the coupling constant $g_i$ is removed),
and studies the choices $\beta_i =$~0.4, 0.5, and 0.7~GeV.  To put
these choices on a common footing, note from Eqs.~(\ref{eq:rhodef})
and (\ref{eq:rhodef2}) that $k_i = \frac 1 2 \sqrt{s - s_{{\rm
      th},i}}$, and that the form in~\cite{Swanson:2014tra} can be
normalized to unity at threshold by multiplying $F^2_i$ by a constant,
$F^2_i = {\rm exp} [-(s-s_{{\rm th},i})/\beta_i^2]$.  Then the
exponential form may be written as
\begin{equation} \label{eq:expform}
F^2_i (s) = {\rm exp} \left( -\frac{\mu_i \rho_i^2}{1-\rho_i^2}
\right) \, ,
\end{equation}
where
\begin{equation} \label{eq:mudef}
\mu_i = \frac{s_{{\rm th},i} R^2}{12} =
\frac{s_{{\rm th},i}}{\beta_i^2} \, ,
\end{equation}
so that $\mu_i = 0.738$ for \cite{Bugg:2008wu}, and $R = 0.6$~fm
corresponds to $\beta_i = 1.14$~GeV\@.  If $k_i$ from
Eq.~(\ref{eq:rhodef}) is identified with the nonrelativistic momentum
$|{\bf k}_i|$, then one may compute the normalized nonrelativistic
matter density $\rho ({\bf r})$ (not to be confused with $\rho_i$) as
its Fourier transform:
\begin{equation} \label{eq:Fourier}
\rho ({\bf r}) = \frac{1}{(2\pi \hbar)^3} \int d^3 {\bf k} \,
e^{-i {\bf k} \cdot {\bf r} / \hbar} F({\bf k}^2) \, ,
\end{equation}
giving in this case a Gaussian form:
\begin{equation}
\rho ({\bf r}) = \left( \frac{3}{4\pi R^2} \right)^{\frac 3  2}
{\rm exp} \left( - \frac{3r^2}{4R^2} \right) \, .
\end{equation}

Inserting Eq.~(\ref{eq:expform}) into Eq.~(\ref{eq:rhodispreln}) leads
to integrals for $\pi_i (s)$ that do not appear to be expressible in
closed form except at special values of $s$.  That the subtraction is
performed at $s=0$, and hence that $\pi_i (0) = 0$, is guaranteed by
noting that $s=0$ in Eq.~(\ref{eq:vdef}) corresponds to $v_i^2 =
-\rho_i^2 = \infty$.  In terms of the Kummer (confluent
hypergeometric) function $U(a,b,z)$, the function $\pi_i(s)$ and its
slope $\pi^\prime_i (s)$, can be computed at a few key values of $s$:
\begin{eqnarray}
\begin{array}{ll}
\pi_i (0) = 0 , & \pi^\prime_i (0) = \frac{1}{2\sqrt{\pi}
s_{{\rm th},i}} \, U \! \left( {\scriptstyle \frac 3 2}, 0, \mu
\right) , \hspace{-0.75em} \\
\pi_i (s_{{\rm th},i}) = \frac{1}{\sqrt{\pi}} \, U \! \left(
{\scriptstyle \frac 1 2} , 0, \mu \right) , &
\pi^\prime_i (s_{{\rm th},i}^-) = \infty , \\
\pi_i (\infty) = -\frac{1}{2\sqrt{\pi}} \, U \!
\left( {\scriptstyle \frac 3 2} , 1, \mu \right) , &
\pi^\prime_i (\infty) = 0 .
\end{array}
& & \nonumber \\ & & \label{eq:mesonvalues}
\end{eqnarray}
These values can also be written as somewhat more complicated
expressions in terms of the modified Bessel functions $K_0$ and
$K_1$.  Note especially the cusp $\pi^\prime (s_{{\rm th},i}^-) =
\infty$ (approaching from $s < s_{{\rm th},i})$, which follows
directly from Eq.~(\ref{eq:rhodispreln}) for any $F^2_i(s)$ that is
smooth at $s=s_{{\rm th},i}$ ($\rho_i = 0$).  The slope
$\pi^\prime (s_{{\rm th},i}^+)$ (approaching from $s >
s_{{\rm th},i}$), on the other hand, is generally finite (and
negative) due to the smoothing effect of the principal-value
prescription in Eq.~(\ref{eq:rhodispreln}).

It is important to note that a form factor $F_i^2 (s)$ exponential in
$s$ cannot truly represent the full physical
amplitude~\cite{Szczepaniak:2015eza} in the entire complex $s$ plane,
since it produces an essential singularity for large $s$ in some
directions.  Such behavior in dispersion relations would lead, for
example, to a violation of causality.  For our purposes, however, one
may suppose that the behavior of $F_i^2 (s)$ remains numerically close
to the exponential form along a substantial portion of the real $s$
axis, but that the exact form contains a functional dependence in $s$
(for example, the power-law fall-off predicted by quark-counting
rules) that restores the proper analytic behavior for all complex
values of $s$.

\subsection{Diquark Form Factors} \label{subsec:diquark}

The discussion of Sec.~\ref{sec:QuarkCount} shows that the tetraquark
form factor $F(s)$ at large $s$ scales as $1/s^3$ due to the exchange
of hard gluons needed to maintain the integrity of the exclusive
tetraquark state.  Indeed, the true scaling is $[\alpha_s (s)/s]^3$;
however, the large-$s$ scaling of $\alpha_s$ is logarithmic and
therefore varies slowly compared to the power-law behavior, and
henceforth is neglected.

The corresponding large-$s$ behavior was not used for the mesonic case
because one expects the four quarks produced at any low or
intermediate $s$ immediately to confine into two hadrons; the only
color interactions between the two mesons are then the ``color van der
Waals'' forces represented by final-state interactions.  In the model
of Ref.~\cite{Brodsky:2014xia}, however, the (colored) diquarks
separate a significant distance before they are forced to hadronize,
and so the fundamental color forces can be expected to remain active
at much lower values of $s$ than for the meson case.  Since the
natural scale at which the high-$s$ scaling should become significant
is given by the diquark mass $\sqrt{s_{{\rm th},i}} = 2m_\delta$, we
model the diquark form factor by
\begin{equation}
F_i (s) = \left( \frac{s_{{\rm th},i}}{s} \right)^3 \, ,
\end{equation}
which of course is properly normalized, $F_i (s_{{\rm th},i}) = 1$.
Using Eqs.~(\ref{eq:rhodef}) and (\ref{eq:Fourier}), the matter
density associated with this form factor is
\begin{equation}
\rho ({\bf r}) = \frac{1}{32\pi r_C^3} \left( 1 + \frac{r}{r_C}
\right) e^{-r/r_C} \, ,
\end{equation}
where $r_C$ is the diquark Compton wavelength $1/m_\delta$.

In this case, the integrals in $\pi_i (s)$ can be performed in closed
form.  First, for $s > s_{{\rm th},i}$, one finds
\begin{eqnarray}
\lefteqn{\pi_i (s) = } & & \nonumber \\
& & \frac{1}{\pi} \bigg\{ -\rho_i \left( 1 - \rho_i^2
\right)^6 \ln \left( \frac{1+\rho_i}{1-\rho_i} \right) +
\frac{2048}{3003} \nonumber \\ & &
- 2\rho_i^2 \! \left( \frac{793}{231} - \frac{667}{63} \rho_i^2 \!
+ \frac{562}{35} \rho_i^4 \! - \frac{66}{5} \rho_i^6 \!
+ \frac{17}{3} \rho_i^8 \! - \rho_i^{10} \! \right) \! \bigg\} \, .
\nonumber \\ & & \label{eq:diquarkrho}
\end{eqnarray}
In the $s < s_{{\rm th},i}$ case, we define [using
Eq.~(\ref{eq:vdef})] the parameter
\begin{equation}
\gamma_i \equiv \frac{s}{s_{{\rm th},i}} = \frac{1}{1 + v_i^2}
\, ,
\end{equation}
which varies from $0 \to 1$ as $s$ increases from $0 \to s_{{\rm
    th},i}$.  One then computes:
\begin{eqnarray}
\lefteqn{ \hspace{-0.7em}
\pi_i (s) = \frac{1}{\pi \gamma_i^6} \bigg\{
 -\sqrt{ \frac{1-\gamma_i}{\gamma_i}}
\left[ \frac{\pi}{2} - \tan^{-1} \left(
\frac{\frac 1 2 - \gamma_i}{\sqrt{\gamma_i (1-\gamma_i)}} \right)
\right] }
\nonumber \\
& & + 2 \left( 1 - \frac 1 3 \gamma_i - \frac{2}{15} \gamma_i^2
- \frac{8}{105} \gamma_i^3 - \frac{16}{315} \gamma_i^4 -
\frac{128}{3465} \gamma_i^5 \right. \nonumber \\
& & \ \ \ \ \ \left.  - \frac{256}{9009} \gamma_i^6 \right)
\bigg\} \, . \label{eq:diquarkbeta}
\end{eqnarray}
These precise forms are not terribly illuminating in their own right,
but they can be used to compute various limits for purposes of
comparison with the mesonic case.  For example,
Eq.~(\ref{eq:diquarkbeta}) appears to have a very strong singularity
at $\gamma_i = 0$ ($s = 0$), but the complicated polynomial in
$\gamma_i$ has precisely the right form to cancel the all terms in the
expansion of the tan$^{-1}$ expression in powers of $\gamma_i$ to the
same order to give $\pi_i (0) = 0$, and furthermore to give a finite
value at $\gamma_i = 1$ ($s = s_{{\rm th},i}$) that matches the one
from Eq.~(\ref{eq:diquarkrho}) at $\rho_i = 0$, but with a derivative
from below that is infinite.  In comparison with
Eq.~(\ref{eq:mesonvalues}),
\begin{eqnarray}
\begin{array}{ll}
\pi_i (0) = 0 , & \pi^\prime_i (0) =
\frac{2048}{45045\pi s_{{\rm th},i}} , \\
\pi_i (s_{{\rm th},i}) = \frac{2048}{3003\pi} , &
\pi^\prime_i (s_{{\rm th},i}^-) = \infty , \\
\pi_i (\infty) = -\frac{512}{9009\pi} , &
\pi^\prime_i (\infty) = 0 .
\end{array}
& &
\label{eq:diquarkvalues}
\end{eqnarray}

\section{Numerical Results} \label{sec:Results}

In this section we present explicit numerical comparisons of the
effects of threshold cusps due to mesons compared to those due to
$\delta$-$\bar \delta$ pairs.  But first, we make a few comments about
how best to present and interpret these effects.

While it is rather suggestive to describe the attractive cusp in
Re~$\! \Pi_i (s)$ at $s = s_{{\rm th}, i}$ as a potential well that
attracts a bare pole to the threshold, it is important not to be
overly seduced by the analogy with a configuration-space potential
well, which always attracts a particle to the vicinity of its minimum.
The intuition one develops for the cusp effect needs to be more
nuanced.  In fact, the entire portion of the cusp over which Re~$\!
\Pi_i (s)$ is positive and appreciable in magnitude provides a source
of attraction for the pole.  For example, even if the bare pole mass
sits exactly at $M_0 = \sqrt{s_{{\rm th}, i}}$, the cusp can drive the
pole mass to a value slightly larger than $\sqrt{s_{{\rm th}, i}}$;
such an effect is visible in the results in Table~II of
Ref.~\cite{Bugg:2008wu}.

The corrected pole mass $M_{\rm pole}$ is the pole of the propagator,
{\it i.e.}, the zero of Eq.~(\ref{eq:Prop}).  As the physical
resonance pole first appears on the second Riemann sheet, one must
take care to choose the sign corresponding to the correct branch in
the self-energy function $\Pi_i (s)$.  The effect of the threshold
cusp on the position of a pole can then be described succinctly in
terms of the dimensionless ratios $M_0/\! \sqrt{s_{{\rm th}, i}}$ and
$M_{\rm pole}/\! \sqrt{s_{{\rm th}, i}}$.  In light of the observation
that the whole profile of the cusp is active in dragging $M_0$ to
$M_{\rm pole}$, one expects that the cusp is most effective in
dragging the pole if either the coupling $g_i$ to the opening
threshold is large, or the cusp function Re~$\! \Pi_i (s)$ is broad in
$s$.  The question of whether the mesonic or diquark threshold cusp is
more effective in pole dragging thus comes down to these two criteria.

\begin{figure}
\begin{center}
\includegraphics[width=\linewidth]{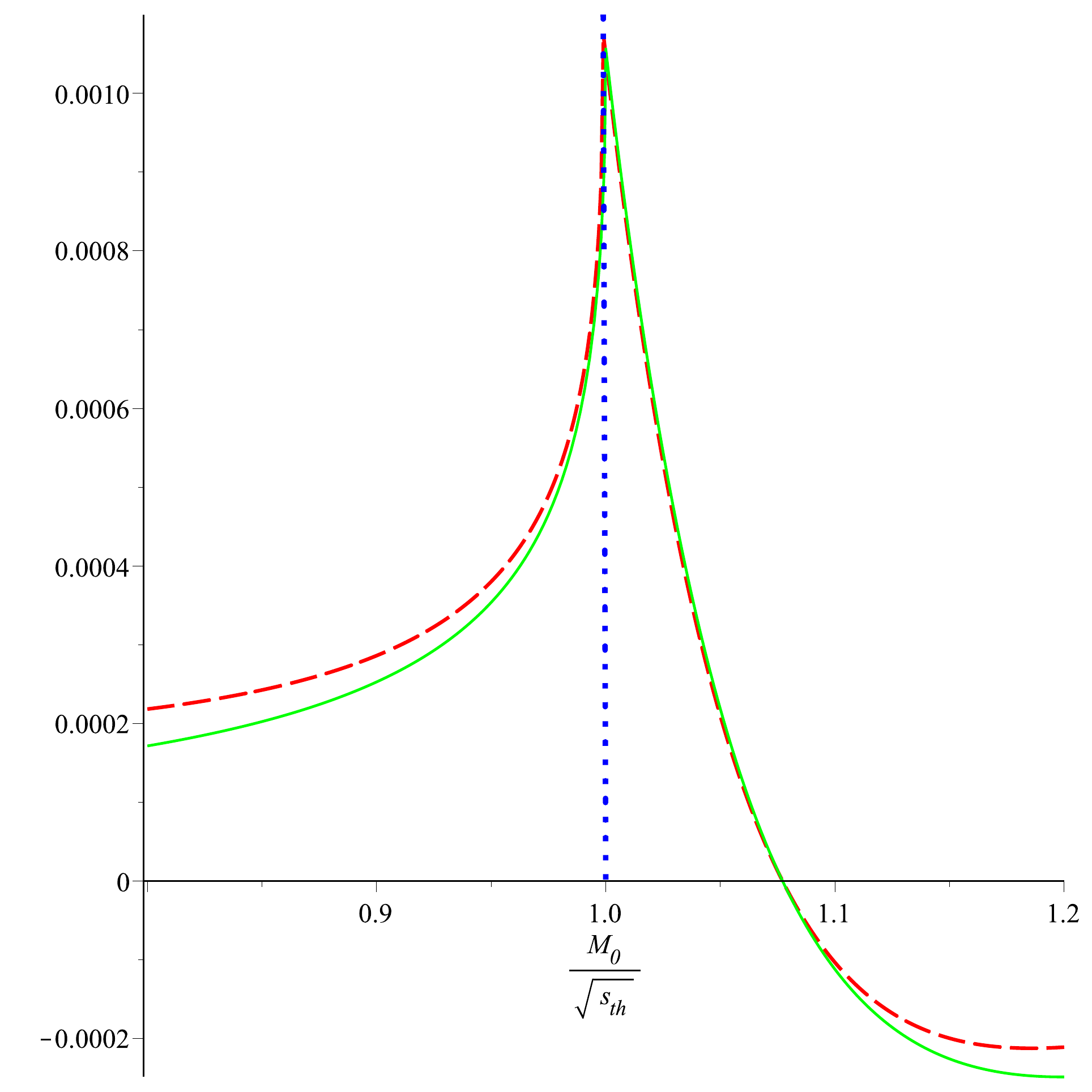}
\caption{A sample threshold cusp (solid, green) using the diquark form
  presented in Sec.~\ref{subsec:diquark}, plotted as a function of $x
  \equiv \sqrt{s/s_{{\rm th},i}}$.  Its peak normalization is chosen
  to match that of the overlaid (dashed, red) pole-dragging effect of
  this cusp, $y \equiv (M_{\rm pole} - M_0) / \! \sqrt{s_{{\rm
        th},i}}$, with $g_i = 0.370$~GeV and $\sqrt{s_{{\rm th},i}} =
  3.872$~GeV, and plotted as a function of a second abscissa $x \equiv
  M_0 / \!  \sqrt{s_{{\rm th},i}}$.  Also plotted is the line $y = 1 -
  x \equiv 1 - M_0 / \! \sqrt{s_{{\rm th},i}}$ (dotted, blue) dividing
  Regions~1 and 2, as defined in the text.
  \label{Fig:Cusp_and_Shift}}
\end{center}
\end{figure}

To exhibit the effectiveness of the cusp numerically, we first
superimpose two plots in Fig.~\ref{Fig:Cusp_and_Shift}.  The first is
a sample (diquark) cusp $\pi_i (s)$ with $g_i = 0.370$~GeV and
$\sqrt{s_{{\rm th},i}} = 3.872$~GeV, which uses the abscissa $x \equiv
s/s_{{\rm th},i}$.  Its peak is normalized at threshold ($x = 1$) to
match that of the second plot, which uses the abscissa $x \equiv M_0/
\! \sqrt{s_{{\rm th},i}}$ and an ordinate that gives a measure of the
effectiveness of the cusp in dragging the pole, chosen to be
\begin{equation} \label{eq:ydef}
y \equiv \frac{M_{\rm pole} - M_0}{\sqrt{s_{{\rm th},i}}} \, ,
\end{equation}
The pole-dragging function $y$ very closely follows the shape of the
cusp, but does not precisely match it.  Qualitatively, this result
means that the cusp does indeed attract the pole over a significant
range of $M_0/ \! \sqrt{s_{{\rm th},i}}$.  To consider the effect more
carefully, we characterize regions of the plot:
\begin{enumerate}
\item $\displaystyle \frac{M_0}{\sqrt{s_{{\rm th},i}}} <
\frac{M_{\rm pole}}{\sqrt{s_{{\rm th},i}}} < 1$: \vspace{1ex} \\
attraction of below-threshold pole toward $s_{{\rm th},i}$;
\item $\displaystyle \frac{M_0}{\sqrt{s_{{\rm th},i}}} <
1 < \frac{M_{\rm pole}}{\sqrt{s_{{\rm th},i}}}$: \vspace{1ex} \\
attraction of below-threshold pole past $s_{{\rm th},i}$;
\item $\displaystyle 1 < \frac{M_0}{\sqrt{s_{{\rm th},i}}} <
\frac{M_{\rm pole}}{\sqrt{s_{{\rm th},i}}}$: \vspace{1ex} \\
repulsion of above-threshold pole from $s_{{\rm th},i}$;
\item $\displaystyle 1 < \frac{M_{\rm pole}}{\sqrt{s_{{\rm th},i}}} <
\frac{M_0}{\sqrt{s_{{\rm th},i}}}$: \vspace{1ex} \\
attraction of above-threshold pole toward $s_{{\rm th},i}$.
\end{enumerate}
The other two possibilities, {\it i.e.}, repulsion of a
below-threshold pole from $s_{{\rm th},i}$ and attraction of an
above-threshold pole past $s_{{\rm th},i}$, do not occur for cusp
functions of the types considered here.  The interesting possibility
(Region~2) mentioned above, of the cusp causing a below-threshold pole
to overshoot threshold, is represented in
Fig.~\ref{Fig:Cusp_and_Shift} as the sliver of the plot between the
line $1 \! -x$ and the vertical line $x = M_0/ \! \sqrt{s_{{\rm
      th},i}} = 1$.  At these input values, the peak of the
pole-dragging function $y$ lies slightly below the peak of the cusp
function (at $x = 1$), but still above the dividing line $1 \! - x$,
so that Region~2 has a finite extent.  Equally interesting is
Region~3, between the lines $x = 1$ and where $y = 0$, in which the
bare pole $M_0$ lies near threshold but is repelled by it nonetheless.
Both of these effects are caused by the continuing attraction (to
larger $s$ values) of the cusp beyond $s_{{\rm th},i}$.
From Fig.~\ref{Fig:Cusp_and_Shift}, one sees that the most effective
dragging of the pole (the maximum of $y$) falls very close to the
dividing line $y = 1 - x$, which is the point where $M_{\rm pole} =
\sqrt{s_{{\rm th},i}}$.

We now turn to a direct comparison between the pole-dragging
efficiency of diquark and mesonic cusps, which are chosen to have the
same value of $s_{{\rm th},i}$.  Of course, the presence of a cusplike
structure not coinciding with a known mesonic threshold could suggest
the presence of a distinct diquark threshold, possibly mixed with a
$\delta$-$\bar \delta$ resonance as discussed here, or even a
heretofore unknown $\bar q q$ resonance; and even in the case that a
meson and diquark threshold coincide or overlap, the shape of their
combined effect would be distinct from that of a meson threshold
alone.  For the diquark form, the only free parameters are the
coupling constant $g_i$ and the bare pole mass as a multiple of
threshold, $M_0/\!  \sqrt{s_{{\rm th},i}}$.  The mesonic cusp, on the
other hand, contains the additional free parameter $\beta_i$
indicative of a typical hadronic scale, which should assume roughly
the same $\approx 0.5$--1.0~GeV value for any hadronic system;
alternately, it can be expressed [Eq.~(\ref{eq:mudef})] as the
dimensionless combination $\mu_i$, which changes from system to system
depending on the value of $s_{{\rm th},i}$.

\begin{figure}
\begin{center}
\includegraphics[width=\linewidth]{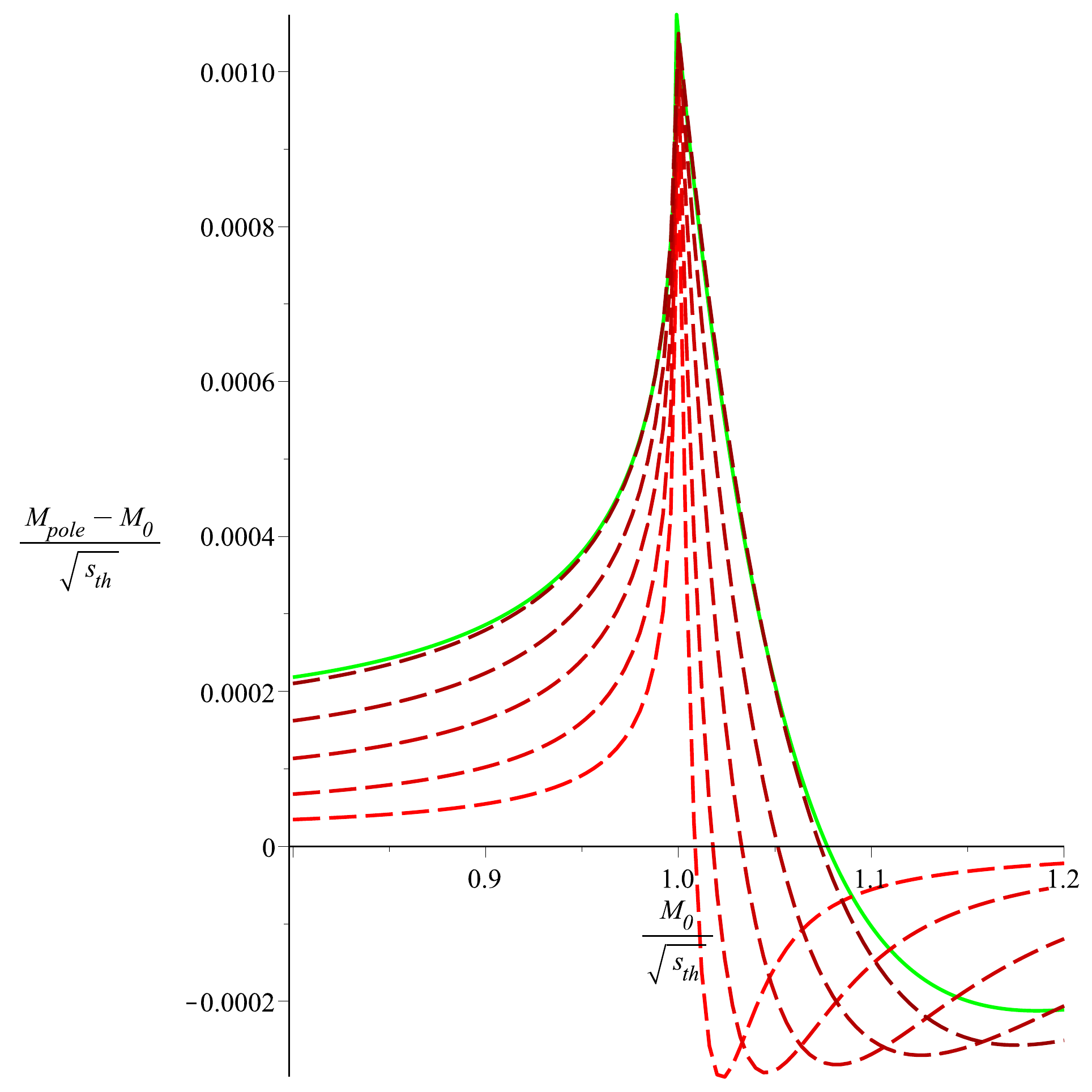}
\caption{Comparison of the effectiveness of pole dragging by cusps
  [$y$ of Eq.~(\ref{eq:ydef})] as a function of $x \equiv M_0/ \!
  \sqrt{s_{{\rm th},i}}$, as created by diquark (solid, green) and
  meson (dashed, red) thresholds.  The values of $\beta_i$ (in GeV)
  corresponding to the mesonic plots, in increasing width of the
  profile (or darkness of the shading), are 0.5, 0.8, 1.1, 1.4, and
  1.7.  The mesonic plots are scaled to the same cusp height, in the
  manner discussed in the text.\label{Fig:Diquark_and_Meson}}
\end{center}
\end{figure}

In Fig.~\ref{Fig:Diquark_and_Meson} we compare the pole-dragging
effectiveness parameter $y$ of Eq.~(\ref{eq:ydef}) as a function of $x
\equiv M_0/ \! \sqrt{s_{{\rm th},i}}$ for the choice $\sqrt{s_{{\rm
      th},i}} = m_{D^0} + m_{D^{*0}} = 3.872$~GeV relevant to the
$X(3872)$.  The diquark pole-dragging plot is identical to that in
Fig.~\ref{Fig:Cusp_and_Shift} (it is the dashed curve there), where it
corresponds to $g_{i, {\rm diquark}} = 0.370$~GeV.  The mesonic
pole-dragging plots are presented for several values of $\beta_i$.
Since, as discussed below, the values of $g_{i, {\rm diquark}}$ are
chosen to scale with $g_{i, {\rm meson}}$ so that the diquark and
mesonic cusp functions have the same height in each case, and since
the absolute height of each mesonic pole-dragging curve changes with
$\beta_i$, one should really view Fig.~\ref{Fig:Diquark_and_Meson} as
a family of plots: a diquark and a mesonic curve for each value of
$\beta_i$.  The diquark plots inherit $\beta_i$ dependence only
through this scaling.  The diquark and mesonic plots give the
indicated value of $y$ only for $\beta_i = 1$~GeV (where $g_{i, {\rm
    meson}} = 0.474$~GeV), and both would be proportionally larger
(smaller) for $\beta_i > 1$~GeV ($< 1$~GeV).  However, for fixed
$\sqrt{s_{{\rm th},i}}$ the diquark curves have a fixed shape, and
hence are scaled to the single (solid) curve in
Fig.~\ref{Fig:Diquark_and_Meson}, while the mesonic curves become
broader as $\beta_i$ grows.  They increase monotonically in width (and
hence in pole-dragging effectiveness) as a function of $\beta_i$,
achieving near-parity with the diquark curve at $\beta_i = 1.7$~GeV\@.
However, such a large value seems inconsistent with that expected from
ordinary hadronic matter; if one limits to, ${\it e.g.}$, values
$\beta_i = 1.1$~GeV or less, then the mesonic curves are much narrower
than the diquark ones scaled to the same height, and hence mesonic
thresholds do not drag resonant poles as effectively as diquark
thresholds.

However, this very interesting result is also dependent upon the
absolute size $s_{{\rm th},i}$ of threshold.  One expects the same
range of $\beta_i$ values to occur due to hadronic confinement
physics, but the mesonic plots depend upon the dimensionless
combination $\mu_i$ of Eq.~(\ref{eq:mudef}).  In the case of the
$K\overline{K}$ threshold studied in Ref.~\cite{Bugg:2008wu},
$\sqrt{s_{{\rm th},i}} = 0.991$~GeV, while for the studies of $Z_b$
states in Ref.~\cite{Swanson:2014tra}, the threshold is $\sqrt{s_{{\rm
      th},i}} = m_B + m_{B^*} = 10.604$~GeV\@.

One must also take into account the chosen size for the coupling
constant $g_i$.  In Fig.~\ref{Fig:Diquark_and_Meson}, the same cusp
function peak value is used for both diquark and meson forms.  Of
course, no diquark pair production coupling constant has ever been
measured experimentally; however, as noted in the Introduction, the
fundamental strength of the color interaction between a $qq$ pair
forming a $\bar {\bf 3}$ is fully half as strong as that between a
$\bar q q$ pair forming a ${\bf 1}$, so it is reasonable to take the
diquark and meson $g_i$ values to be comparable in size.  In the plot
captions, $g_{i, {\rm meson}}$ is the given numerical value, and
$g_{i, {\rm diquark}}$ is derived from normalizing the peak of its
cusp function to match that of the mesonic form; this ratio is
computed using the values of $\pi_i(s_{{\rm th},i})$ in
Eqs.~(\ref{eq:mesonvalues}) and (\ref{eq:diquarkvalues}).

Furthermore, the coupling $g_i$ corresponds to the creation of hadrons
containing heavy quarks of species $Q$ ($s$ for $K\overline{K}$, $c$
for $D\bar D^*$, $b$ for $B\bar B^*$), and therefore is proportional
to the decay constant $f_Q$, which is known from heavy-quark effective
theory to scale in terms of the heavy-quark mass $m_Q$ as $1/\!
\sqrt{m_Q}$.  In turn, the relevant thresholds scale as $\sqrt{s_{{\rm
      th},i}} \sim m_Q$, so that $g_i \sim (s_{{\rm th}, i})^{-1/4}$.
The means by which all given coupling constants $g_{i, {\rm meson}}$
are given is therefore
\begin{equation} \label{eq:HQET_Scaling}
g_i^2 = g_{K\bar{K}}^2 \sqrt{ \frac{s_{{\rm th}, K\bar{K}}}
{s_{{\rm th}, i}}} \, ,
\end{equation}
with $g_{K\bar{K}}^2 = 0.875$~GeV$^2$, an example given in
Ref.~\cite{Bugg:2008wu}.  Clearly, one may dispute treating the $s$
quark as heavy, the accuracy of using $\sqrt{s_{{\rm th},i}}$ in place
of $m_Q$, or indeed the scaling of diquark and meson cusp functions to
have the same peak height.  However, all of these issues may be
adjusted to match one's prejudices by including appropriately chosen
$O(1)$ correction coefficients.  Our prescriptions are designed to
make direct comparisons between the forms as clear as possible.

\begin{figure}
\begin{center}
\includegraphics[width=\linewidth]{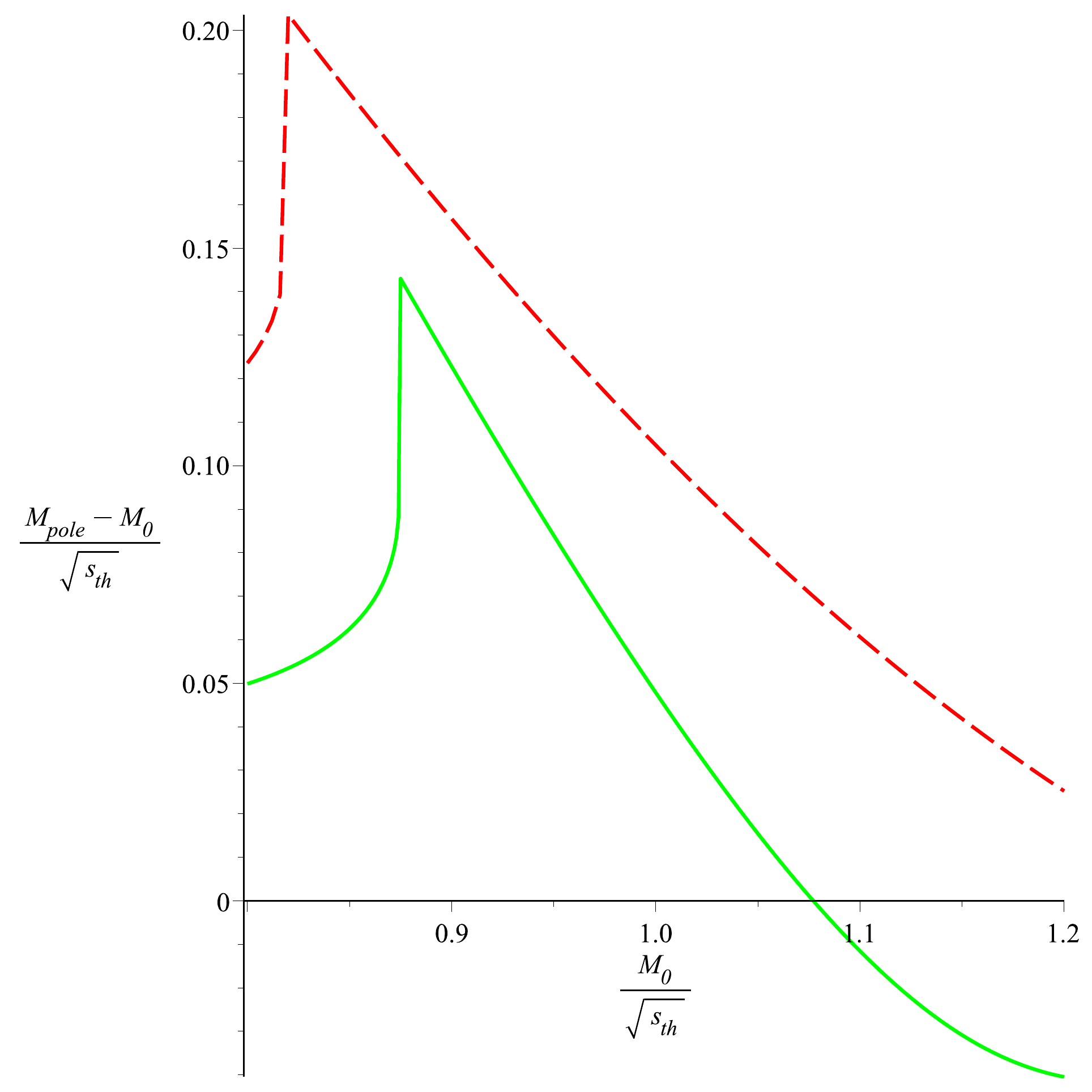}
\caption{Comparison of the effectiveness of pole dragging by cusps
  [$y$ of Eq.~(\ref{eq:ydef})] as a function of $x \equiv M_0/ \!
  \sqrt{s_{{\rm th},i}}$, from the diquark cusp (solid, green) and
  from the mesonic cusp (dashed, red) chosen to have $\beta_i =
  1.0$~GeV\@.  Here, $\sqrt{s_{{\rm th},i}} = 0.991$~GeV, and $g_{i,
    {\rm diquark}} = 1.348$~GeV, while $g_{i, {\rm meson}} =
  0.935$~GeV to give the diquark and meson cusp functions the same
  height.
  \label{Fig:KK}}
\end{center}
\end{figure}

\begin{figure}
\begin{center}
\includegraphics[width=\linewidth]{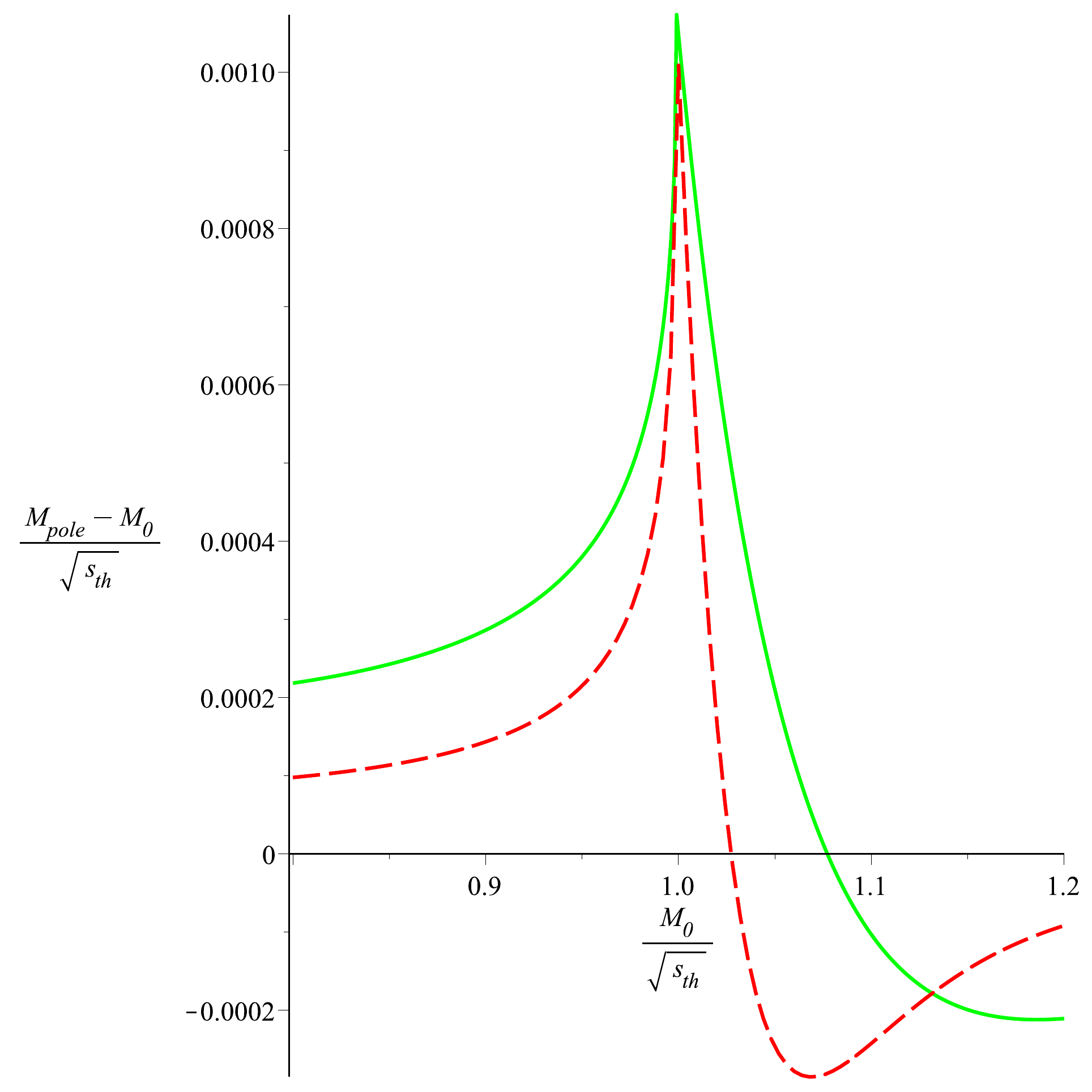}
  \caption{As in Fig.~\ref{Fig:KK}, except with $\sqrt{s_{{\rm th},i}}
  = 3.872$~GeV, and $g_{i, {\rm diquark}} = 0.370$~GeV, while $g_{i,
  {\rm meson}} = 0.474$~GeV to give the diquark and meson cusp
  functions the same height.  This case can be compared directly to
  that with $\beta = 1$~GeV in Fig.~\ref{Fig:Diquark_and_Meson}. 
  \label{Fig:DD}}
\end{center}
\end{figure}

\begin{figure}
\begin{center}
\includegraphics[width=\linewidth]{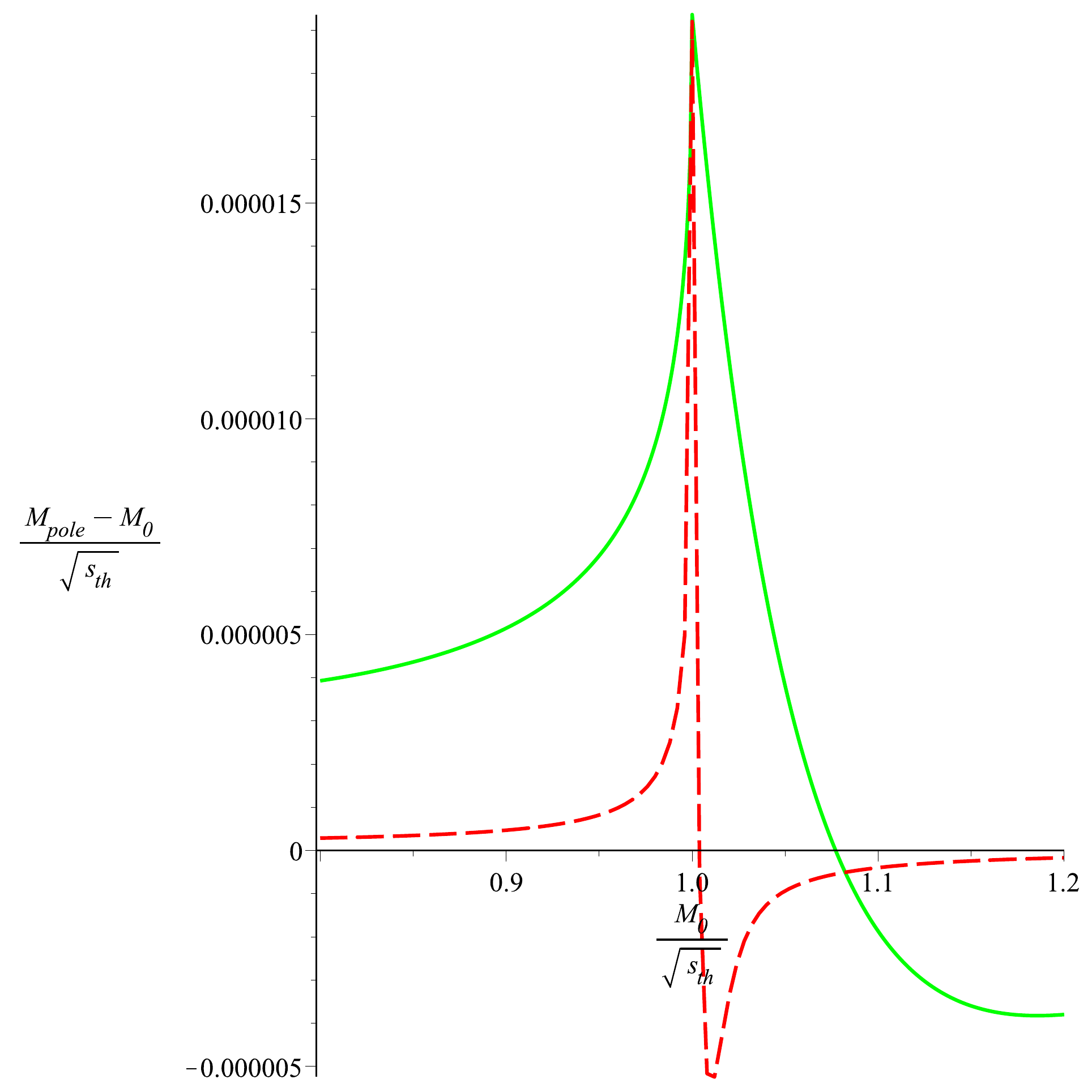}
\caption{As in Fig.~\ref{Fig:KK}, except with $\sqrt{s_{{\rm th},i}} =
  10.604$~GeV, and $g_{i, {\rm diquark}} = 0.136$~GeV, while $g_{i,
    {\rm meson}} = 0.286$~GeV to give the diquark and meson cusp
  functions the same height.
\label{Fig:BB}}
\end{center}
\end{figure}

In Figs.~\ref{Fig:KK}, \ref{Fig:DD}, and \ref{Fig:BB} we compare the
pole-dragging effectiveness parameter $y$ obtained for diquark
thresholds, alongside the mesonic form with the $K\overline{K}$,
$D\bar D^{*}$, and $B\bar B^*$ thresholds, respectively, each with
$\beta_i = 1.0$~GeV.  One sees that the $K\overline{K}$ mesonic curve
is much wider than the diquark curve, while the $D\bar D^{*}$ and
$B\bar B^*$ mesonic curves are much narrower.  We infer that the
presence of a $\delta$-$\bar \delta$ threshold is more effective in
dragging a resonant pole for heavy-quark systems, and the two-meson
threshold is more effective in dragging a resonant pole for
light-quark systems.

But several other interesting results follow from Figs.~\ref{Fig:KK},
\ref{Fig:DD}, and \ref{Fig:BB}. First, the maximum height of the
pole-dragging function $y$ is much larger in the $K\overline{K}$ case
than in the $D\bar D^*$ or $B\bar B^*$ cases, for either diquark or
mesonic forms.  Even taking into account the larger value of
$\sqrt{s_{{\rm th},i}}$ for the latter cases, the absolute size of the
maximal pole dragging is larger for the lighter cases; to wit, the
numbers are about 200~MeV, 4.3~MeV, and 0.2~MeV, respectively.  Some,
but not all, of this decreased effectiveness can be attributed to the
decrease of $g_i$ via heavy-quark scaling given by
Eq.~(\ref{eq:HQET_Scaling}).  Indeed, for a fixed value of $s_{{\rm
    th},i}$, the relative size of the pole-dragging effect $y$ is
found empirically to scale approximately as $g_i^2$; clearly, the
magnitude of $\sqrt{s_{{\rm th},i}}$ matters as well.  In a related
effect, the location of the peak in $y$ travels a greater distance
from threshold for lighter systems, and it is somewhat larger for the
mesonic than the diquark effect, even when the cusp functions
themselves have the same peak height.  However, one should not
conclude from these facts that the cusp effect is intrinsically less
effective for heavy-quark systems; indeed, a number of the mass
splittings between heavy hadrons scale as $1/m_Q$, so a full analysis
would require one to take into account values of $M_0$ that lie
naturally closer to $\sqrt{s_{{\rm th},i}}$.

\section{Discussion and Conclusions} \label{sec:Concl}

We have performed the first analysis of the cusp effect due to the
opening of diquark-antidiquark thresholds by using constituent
counting rules to model their production form factor.  We directly
compared our results to those obtained from employing a
frequently-used phenomenologically-based meson form factor, and found
that the magnitude of the pole-dragging effect is greater in both
relative and absolute terms for lighter systems ($K\overline K$ vs.\
$D\bar D^*$ or $B\bar B^*$).  We also found that the effect due to
mesonic cusps is larger than that due to diquark cusps for the
$K\overline K$ threshold, while the diquark cusp effect is stronger
for $D\bar D^*$ or $B\bar B^*$ thresholds.

This calculation is of course an idealization, in that only one
threshold is present in each example; in reality, several thresholds,
each with its own strength and contributing with different signs (due
to the different Riemann sheets) are simultaneously present and all
contribute to the total effect.  Furthermore, this calculation
assumes, for maximum clarity of comparison, that the coupling
constants for the diquark and mesonic cases give cusp functions of the
same height, and that the coupling constants scale with the thresholds
according to expectations from heavy-quark effective theory.  Any of
these approximations can be relaxed in a more detailed analysis.

The central conclusion, however, is that if $\delta$-$\bar \delta$
states exist in the spectrum of QCD, then the opening of their
production thresholds produce measurable shifts in the masses of
resonances, which must be taken into account in precisely the same way
as shifts appearing due to the opening of meson production thresholds.
The cusp effect appears to promise a rich source of new physical
effects.

\begin{acknowledgments}
  This work was supported by the National Science Foundation under
  Grant Nos.\ PHY-1068286 and PHY-1403891.  R.F.L.\ thanks S.~Brodsky
  and D.~Bugg for encouragement and useful discussions, and F.-K.~Guo
  and C.~Hanhart for interesting dialogue and insights.
\end{acknowledgments}

\appendix*

\section{Expressions for Unequal Masses}

In the case $m_{1,i} \neq m_{2,i}$, let us define
\begin{eqnarray}
m        & \equiv & \frac 1 2 ( m_{1,i} + m_{2,i} ) \, , \nonumber \\
\delta   & \equiv & \frac 1 2 ( m_{1,i} - m_{2,i} ) \, , \nonumber \\
\epsilon & \equiv & \frac{\delta^2}{m^2} \, ,
\end{eqnarray}
and henceforth suppress the index $i$.  Note that $s_{{\rm th},i} =
4m^2$.  The kinematical variables are, in analogue to
Eqs.~(\ref{eq:rhodef}) and (\ref{eq:rhodef2}),
\begin{equation}
\rho = \frac{2k}{\sqrt{s}} = \sqrt{ \left( 1 - \frac{4m^2}{s} \right)
\left( 1 - \frac{4\delta^2}{s} \right) } \, ,
\end{equation}
and its inverse reads
\begin{equation}
s = \frac{2m^2}{1-\rho^2} (1 + \epsilon) (1 + h) \, ,
\end{equation}
written in terms of the auxiliary variable
\begin{equation}
h \equiv \sqrt{ 1 - \frac{4\epsilon (1-\rho^2) }{(1+\epsilon)^2} } \, ,
\end{equation}
which equals 1 in the equal-mass case $\epsilon = 0$.  Analogous
primed forms hold for $s \to s^\prime$.

Then the generalization of Eq.~(\ref{eq:rhodispreln}) becomes
\begin{equation}
\pi (s) = \frac{1}{\pi} \, {\rm P} \! \int_0^1 d\rho^\prime
\frac{\rho^{\prime \, 2}}{h^\prime} \frac{F^2 (s^\prime)}{\left(
\frac{1-\rho^2}{1+h} - \frac{1-\rho^{\prime \, 2}}{1+h^\prime}
\right)} \, ,
\end{equation}
and that of Eq.~(\ref{eq:rhodispreln2}) is
\begin{eqnarray}
\lefteqn{\pi (s) =} & & \nonumber \\
& & -\frac{1}{\pi} \int_0^1 d\rho^\prime \left[ F^2 (s^\prime) +
\frac{4m^2  \rho^{\prime \, 2} (1+\epsilon) (1+h^\prime)}
{(1-\rho^{\prime \, 2})^2 h^\prime} \frac{dF^2 (s^\prime)}{ds^\prime}
\right] \nonumber \\ & & \ \ \times \ln \left| (1-\rho^2) -
(1-\rho^{\prime \, 2}) \frac{1+h \, }{1+h^\prime} \right| \, .
\end{eqnarray}



\begin{thebibliography}{99}
%
\bibitem{Aaij:2014jqa} 
  R.~Aaij {\it et al.}  [LHCb Collaboration],
  Phys.\ Rev.\ Lett.\ {\bf 112}, 222002 (2014)
  [arXiv:1404.1903 [hep-ex]].
%
\bibitem{Choi:2003ue} 
  S.K.~Choi {\it et al.}  [Belle Collaboration],
  Phys.\ Rev.\ Lett.\ {\bf 91}, 262001 (2003)
  [hep-ex/0309032].
%
\bibitem{Brambilla:2014aaa} 
  N.~Brambilla, S.~Eidelman, P.~Foka, S.~Gardner, A.S.~Kron\-feld,
  M.G.~Alford, R.~Alkofer, and M.~Butensch\"{o}n {\it et al.},
  [arXiv:1404.3723 [hep-ph]].
%
\bibitem{Esposito:2013ada} 
  A.~Esposito, F.~Piccinini, A.~Pilloni and A.D.~Polosa,
  J.\ Mod.\ Phys.\ {\bf 4}, 1569 (2013)
  [arXiv:1305.0527 [hep-ph]].
%
\bibitem{Guerrieri:2014gfa} 
  A.L.~Guerrieri, F.~Piccinini, A.~Pilloni and A.D.~Polosa,
  Phys.\ Rev.\ D {\bf 90}, 034003 (2014)
  [arXiv:1405.7929 [hep-ph]].
%
\bibitem{Artoisenet:2009wk} 
  P.~Artoisenet and E.~Braaten,
  Phys.\ Rev.\ D {\bf 81}, 114018 (2010)
  [arXiv:0911.2016 [hep-ph]].
%
\bibitem{Artoisenet:2010uu} 
  P.~Artoisenet and E.~Braaten,
  Phys.\ Rev.\ D {\bf 83}, 014019 (2011)
  [arXiv:1007.2868 [hep-ph]].
%
\bibitem{Voloshin:2007dx} 
  M.B.~Voloshin,
  Prog.\ Part.\ Nucl.\ Phys.\ {\bf 61}, 455 (2008)
  [arXiv:0711.4556 [hep-ph]].
%
\bibitem{Maiani:2004vq} 
  L.~Maiani, F.~Piccinini, A.D.~Polosa, and V.~Riquer,
  Phys.\ Rev.\ D {\bf 71}, 014028 (2005)
  [hep-ph/0412098].
%
\bibitem{Maiani:2014aja} 
  L.~Maiani, F.~Piccinini, A.D.~Polosa and V.~Riquer,
  arXiv:1405.1551 [hep-ph].
%
\bibitem{Brodsky:2014xia} 
  S.J.~Brodsky, D.S.~Hwang and R.F.~Lebed,
  Phys.\ Rev.\ Lett.\ {\bf 113}, 112001 (2014)
  [arXiv:1406.7281 [hep-ph]].
%
\bibitem{Yuan:2014rta} 
  C.-Z.~Yuan,
  Int.\ J.\ Mod.\ Phys.\ A {\bf 29}, 1430046 (2014)
  [arXiv:1404.7768 [hep-ex]].
%
\bibitem{Matveev:1973ra} 
  V.A.~Matveev, R.M.~Muradian and A.N.~Tavkhelidze,
  Lett.\ Nuovo Cim.\ {\bf 7}, 719 (1973).
%
\bibitem{Brodsky:1973kr} 
  S.J.~Brodsky and G.R.~Farrar,
  Phys.\ Rev.\ Lett.\ {\bf 31}, 1153 (1973).
%
\bibitem{Brodsky:1974vy} 
  S.J.~Brodsky and G.R.~Farrar,
  Phys.\ Rev.\ D {\bf 11}, 1309 (1975).
%
\bibitem{Farrar:1979aw} 
  G.R.~Farrar and D.R.~Jackson,
  Phys.\ Rev.\ Lett.\ {\bf 43}, 246 (1979).
%
\bibitem{Efremov:1978rn} 
  A.V.~Efremov and A.V.~Radyushkin,
  Theor.\ Math.\ Phys.\ {\bf 42}, 97 (1980)
  [Teor.\ Mat.\ Fiz.\ {\bf 42}, 147 (1980)].
%
\bibitem{Duncan:1979hi} 
  A.~Duncan and A.H.~Mueller,
  Phys.\ Rev.\ D {\bf 21}, 1636 (1980).
%
\bibitem{Duncan:1980qd} 
  A.~Duncan and A.H.~Mueller,
  Phys.\ Lett.\ B {\bf 93}, 119 (1980).
%
\bibitem{Lepage:1980fj} 
  G.P.~Lepage and S.J.~Brodsky,
  Phys.\ Rev.\ D {\bf 22}, 2157 (1980).
%
\bibitem{Sivers:1982wk} 
  D.W.~Sivers,
  Ann.\ Rev.\ Nucl.\ Part.\ Sci.\ {\bf 32}, 149 (1982).
%
\bibitem{Mueller:1981sg} 
  A.H.~Mueller,
  Phys.\ Rept.\ {\bf 73}, 237 (1981).
%
\bibitem{Brodsky:1989pv} 
  S.J.~Brodsky and G.P.~Lepage,
  Adv.\ Ser.\ Direct.\ High Energy Phys.\ {\bf 5}, 93 (1989).
%
\bibitem{Brodsky:2015XXX}
  S.J.~Brodsky and R.F.~Lebed,
  in preparation.
%
\bibitem{Tornqvist:1995kr} 
  N.A.~T\"{o}rnqvist,
  Z.\ Phys.\ C {\bf 68}, 647 (1995)
  [hep-ph/9504372].
%
\bibitem{Bugg:2008wu} 
  D.V.~Bugg,
  J.\ Phys.\ G {\bf 35}, 075005 (2008)
  [arXiv:0802.0934 [hep-ph]].
%
\bibitem{Guo:2014iya} 
  F.K.~Guo, C.~Hanhart, Q.~Wang, and Q.~Zhao,
  arXiv:1411.5584 [hep-ph].
%
\bibitem{Bugg:2011jr} 
  D.V.~Bugg,
  Europhys.\ Lett.\ {\bf 96}, 11002 (2011)
  [arXiv:1105.5492 [hep-ph]].
%
\bibitem{Swanson:2014tra} 
  E.S.~Swanson,
  Phys.\ Rev.\ D {\bf 91}, 034009 (2015)
  [arXiv:1409.3291 [hep-ph]].
%
\bibitem{Peskin:1995ev} 
  M.E.~Peskin and D.V.~Schroeder,
  {\it An Introduction to quantum field theory},
  (Perseus, New York, USA, 1995).
%
\bibitem{Georgi:1985kw} 
  H.~Georgi,
  {\it Weak Interactions and Modern Particle Theory},
  (Dover Publications, New York, USA, 2009).
%
\bibitem{Hwang:2001th} 
  C.-W.~Hwang,
  Eur.\ Phys.\ J.\ C {\bf 23}, 585 (2002)
  [hep-ph/0112237].
%
\bibitem{Polchinski:2001tt} 
  J.~Polchinski and M.J.~Strassler,
  Phys.\ Rev.\ Lett.\ {\bf 88}, 031601 (2002)
  [hep-th/0109174].
%
\bibitem{Kawamura:2013iia} 
  H.~Kawamura, S.~Kumano and T.~Sekihara,
  Phys.\ Rev.\ D {\bf 88}, 034010 (2013)
  [arXiv:1307.0362 [hep-ph]].
%
\bibitem{Kawamura:2013yya} 
  H.~Kawamura, S.~Kumano and T.~Sekihara,
  JPS Conf.\ Proc.\ {\bf 1}, 013043 (2014)
  [arXiv:1312.1749 [hep-ph]].
%
%
\bibitem{Lepage:1979zb} 
  G.P.~Lepage and S.J.~Brodsky,
  Phys.\ Lett.\ B {\bf 87}, 359 (1979).
%
\bibitem{Efremov:1979qk} 
  A.V.~Efremov and A.V.~Radyushkin,
  Phys.\ Lett.\ B {\bf 94}, 245 (1980).
%
\bibitem{Sudakov:1954sw} 
  V.V.~Sudakov,
  Sov.\ Phys.\ JETP {\bf 3}, 65 (1956)
  [Zh.\ Eksp.\ Teor.\ Fiz.\ {\bf 30}, 87 (1956)].
%
\bibitem{Cornwall:1975ty} 
  J.M.~Cornwall and G.~Tiktopoulos,
  Phys.\ Rev.\ D {\bf 13}, 3370 (1976).
%
\bibitem{Sen:1982bt} 
  A.~Sen,
  Phys.\ Rev.\ D {\bf 28}, 860 (1983).
%
\bibitem{Landshoff:1974ew} 
  P.V.~Landshoff,
  Phys.\ Rev.\ D {\bf 10}, 1024 (1974).
%
\bibitem{Li:1992nu} 
  H.-N.~Li and G.F.~Sterman,
  Nucl.\ Phys.\ B {\bf 381}, 129 (1992).
%
\bibitem{Szczepaniak:2015eza} 
  A.P.~Szczepaniak,
  arXiv:1501.01691 [hep-ph].
%
\end{thebibliography}
\end{document}